\input harvmac
\input epsf
\epsfverbosetrue

\def\p{\partial}
\def\ap{\alpha'}

\def\b0{\bar{0}}
\def\b4{\bar{4}}
\def\lmb{\lambda}
\Title{EFI-98-25}{\vbox{\centerline{Evidence for Large N Phase Transition
in ${\cal N}=4$}
\vskip12pt
\centerline{Super Yang-Mills Theory at Finite Temperature}
}}
\vskip20pt
\centerline{Miao Li}
\bigskip
\centerline{\it Enrico Fermi Institute}
\centerline{\it University of Chicago}
\centerline{\it 5640 Ellis Avenue, Chicago, IL 60637, USA}

\bigskip

The AdS/CFT correspondence provides valuable constraints on the possible
exact form of various physical quantities in the ${\cal N}=4$ super 
Yang-Mills theory in the large N limit. 
We examine the free energy as the 
expansions in a small as well as in a large 't Hooft parameter $\lambda$.
We argue that it is impossible to smoothly extrapolate from the
weak coupling regime to the strong coupling regime, thus there must
exist a large N phase transition in $\lmb$ at a finite temperature. 
We also argue that there is no world-sheet instanton in the background 
of the Euclidean anti-de Sitter black hole.

\Date{July 1998}

\nref\jm{J. Maldacena, hep-th/9711200. }
\nref\gkp{S. S. Gubser, I. R. Klebanov and A. M. Polyakov, hep-th/9802109.}
\nref\ew{E. Witten, hep-th/9802150. }
\nref\ewi{E. Witten, hep-th/9803131.}
\nref\jmw{J. Maldacena, hep-th/9803002; S.-J. Rey and   Yee, 
hep-th/9803001.}
\nref\dl{M. R. Douglas and M. Li, to appear.}
\nref\bthree{S. S. Gubser, I. R. Klebanov and A. W. Peet, hep-th/9602135.}
\nref\gkt{S. S. Gubser, I. R. Klebanov and A. A. Tseytlin, hep-th/9805156.}
\nref\bg{T. Banks and M. B. Green, hep-th/9804170.}
\nref\kap{J. I. Kapusta, ``Finite temperature field theory'', Cambridge
University Press (1989).}
\nref\beta{M. T. Grisaru and D. Zanon, Phys. Lett. B177 (1986) 347;
M. D. Freeman, C. N. Pope, M. F. Sohnius and K. S. Stelle, Phys. Lett.
B178 (1986) 199; Q.-H. Park and D. Zanon, Phys. Rev. D35 (1987) 4038.}
\nref\gw{D. Gross and E. Witten, Nucl. Phys. B277 (1986) 1.}
\nref\oldm{E. Brezin and S. R. Wadia, ``The large N expansion in 
quantum field theory and statistical physics: from spin system to
two-dimensional gravity'', World Scientific (1993).}
\nref\plan{J. Koplik, A. Neveu and S. Nussinov, Nucl. Phys.
B123 (1977); W. T. Tuttle, Can. J. Math. 14 (1962) 21.}
\nref\gth{G. 't Hooft, in ``Progress in gauge field theory'', eds.
G. 't Hooft et al. Plenum Press (1984).}
\nref\elect{A. Brandhuber, N. Itzhaki, J. Sonnenschein and
S. Yankielowicz, hep-th/9803137; S.-J. Rey, S. Theisen and J.-T Yee, 
hep-th/9803135.}
\nref\miao{M. Li, hep-th/9803252; hep-th/9804175.}
\nref\go{D. J. Gross and H. Ooguri, hep-th/9805129.}
\nref\gol{J. Greensite and P. Olesen, hep-th/9806235.}

The AdS/CFT correspondence conjectured by Maldacena \jm\ has enabled us 
to study, for instance, ${\cal N}=4$ super Yang-Mills theory \refs{\gkp,
\ew} as well as nonsupersymmetric Yang-Mills theories \ewi\ in the 
large N and the large 't Hooft parameter $\lambda=g^2_{YM}N$ limit.
Many qualitative verifications of this conjecture,  and in some cases
quantitative predictions have been obtained. 

One generic feature of those quantitative predictions is that physical
quantities as functions of $\lambda$ in the large N limit typically
have $\lambda=\infty$ as a branch point. Examples are anomalous
scaling dimensions \refs{\gkp, \ew}, the rectangular Wilson loops
\jmw\ and the leading correction to the free energy \gkt. For a finite
theory such as the ${\cal N}=4$ SYM, it was argued that $\lambda=0$
should be a regular point order by order in the large N expansion
\bg. As in the old matrix models \oldm, one hopes that a physical 
quantity
in the large N limit should have a Taylor expansion in $\lambda$,
and the convergent radius is finite and probably universal.
This radius is largely determined
by the exponential growth of the number of the planar Feynman diagrams 
in the perturbation theory \refs{\plan, \gth}. 
An ansatz for the Wilson loop is proposed in \dl\ and according to
this ansatz $\lambda_c=-2\pi^2$ is a singular point. (This can be
modified by a factor of a rational number.) It is actually
the second branch point as required by the fact that $\lambda=\infty$
is a branch point. Thus we expect that $2\pi^2$ or something similar
is the universal convergent radius. 

We generalize that analysis to the free energy in this note.
We show that an attempt to extrapolate from the weak coupling 
regime to the strong coupling regime runs into contradiction.
Therefore there exists a phase transition at a positive critical
value of $\lmb$. We also discuss the possibility of the world-sheet
instantons in the Euclidean anti-de Sitter black hole background.

\newsec{Large and small $\lmb$ expansions}

In a perturbative calculation of the free energy, the growth of the
number of Feynman diagrams should be the same as in calculating a zero
temperature physical quantity. The only modification to be made is to
replace the usual propagators by the thermal propagators. 
This modification makes the perturbative behavior of a finite
temperature quantity drastically different from its zero temperature
counterpart. The major cause is that an electric mass as well as
a mass for scalars are generated at one-loop level. This mass scale is
proportional to $\sqrt{\lambda}T$. Thus, the perturbative expansion
of a physical quantity is no longer analytic in $\lambda$, but in
$\sqrt{\lambda}$. For instance, the free energy, beyond the
two loop result proportional to $\lambda$, will have a term 
$\lambda^{3/2}$ due to the summation of the so-called ring diagrams
\kap. Each ring diagram of a bosonic field is infrared divergent,
but the summation is finite. The nonanalytic terms generated this
way can not be cancelled by exploiting supersymmetry, since there is no 
such infrared divergent diagram for a fermionic field, due to the fact
that a mass for a fermion is already generated at the tree level
due to the anti-periodic boundary condition. Thus, we expect
the perturbation series of the free energy take the following
schematic form
\eqn\pert{F=\sum_{n=0}^\infty\lambda^n +\sum_{n=0}^\infty
\lmb^{{3\over 2}+n},}
powers of larger half integers can be generated, for instance, by
higher order terms in the electric mass.
Note that we have assumed terms such as $\lmb^2\ln\lmb$ occurring
in QCD are absent here. Such terms, if exist, can be generated in
our scheme presented in the next section.

The free energy of the large N SYM was calculated in \bthree\ in
the $\lambda=0$ limit, and found to be 
$$F=-{\pi^2\over 6}N^2V_3T^4,$$
where $V_3$ is the three volume. However the black hole 
thermodynamics predicts 
$$F=-{\pi^2\over 8}N^2V_3T^4.$$
The mismatch is a factor $3/4$. The black hole result is the
result for $\lambda=\infty$, thus there is no contradiction here.
If the exact formula is $F=-\pi^2/6 N^2V_3T^4F(\lambda)$, then
$F(0)=1$ and $F(\infty)=3/4$. 
A two-loop result is not known yet. On general grounds, one expects
to have
\eqn\twol{F_{two-loop}=a{\lambda\over \pi^2},}
where $a$ is a rational number. 

Using the stringy correction to the effective action \refs{
\beta, \gw}, it was found that in the large $\lambda$ limit
the next to leading order correction is \gkt
\eqn\subl{{45\over 32}\zeta (3)(2\lambda)^{-3/2}.}
It has a positive sign.
We rewrite, for reason becomes obvious shortly, 
\eqn\defi{F(\lambda) =3/4 +G(\lambda),}
then $G(\lambda)$ behaves as $\lambda^{-3/2}$
in the large $\lambda$ limit, and its value at $\lambda=0$
is $1/4$. As discussed above, we expect that it has a Taylor
expansion for small $\lambda$, corresponding to the usual
perturbative expansion. For large $\lambda$, we expect to
have, schematically
\eqn\large{G(\lambda)=\sum_{n\ge 0}\lambda^{-{3\over 2}-n}
+\sum_{n\ge 0}\lambda^{-2-n},}
where the first sum comes from corrections of form $(\ap)^{3+2n}$,
and the second sum comes from corrections of form $(\ap )^{4+2n}$.
Throughout this paper $n$ will be used to denote a natural number,
except otherwise explicitly stated.

One of the ways to see the origin the coefficient $\zeta (3)$ of
the term $\lmb^{-3/2}$ is through an inspection of the 
Virasoro-Shapiro amplitude \gw
\eqn\vsa{A={1\over stu}{\Gamma (1-\ap s)\Gamma(1-\ap t)\Gamma (1-\ap
u)\over \Gamma (1+\ap s)\Gamma(1+\ap t)\Gamma (1+\ap u)}.}
Expanding the above formula in $\ap$ is equivalent to expanding
it in terms of the Mandelstam variables. This gets translated into
the effective action as expansion with number of derivatives.
At the order $(\ap)^3$, the coefficient $d^3\ln \Gamma (x=1)/dx^3$
appears, which is the origin of the factor $\zeta (3)$. 
At the order $(\ap)^{n+3}$, many such derivatives will appear,
and the highest one is just $d^{n+3}\ln\Gamma (x=1)/dx^{n+3}=
(-1)^{n+3}\Gamma (n+3)\zeta (n+3)$. Thus, we expect that the
coefficient of the term $\lmb^{-n/2-3/2}$ which corresponds
to the order $(\ap)^{n+3}$ take the form
\eqn\rieman{\zeta (n+3)+\zeta (3)\zeta (n) +\dots,}
where the general term is a product of 
zeta's whose arguments sum to $n+3$, and each argument is an odd integer. 

For a zero temperature physical quantity, such as the interaction
strength of a heavy quark-anti-quark pair, there are two branch
points on the complex $\lmb$ plane, one is located at infinity,
another is $\lmb_c<0$. New features appear for a finite temperature
quantity, even if we assume that there is no phase transition, so
that the quantity is described by the same function for both small
$\lmb$ and large $\lmb$. We see from the above analysis, that there are
already two branch points, one is infinity, another is $\lmb =0$.
The latter is new, due to infrared divergences at a finite
temperature. Now that it is perfectly possible that there exists
a branch point at a positive $\lmb$, thus signaling a phase transition.
There is no need to connect the branch point at $\lmb =0$ and the
one at $\lmb =\infty$, if there is a phase transition, since 
the free energy is by no means governed by the same function on
the whole complex plane.

\newsec{Failure of interpolation}

Assuming that $G(\lmb )$ is a smooth function on the whole positive
axis, then the following Mellin transform
\eqn\metra{G^*(s)=\int_0^\infty G(\lmb )\lmb^{s-1}d\lmb }
exists for $0<\Re s<3/2$, owing to the decay behavior $G(\lmb )\sim
\lmb^{-3/2}$ for large $\lmb$. Note that the function $F(\lmb )$
does not have a Mellin transform. Given $G^*(s)$ defined for a range
of $s$, it can be analytically extended to the whole complex plane of $s$.
Since $G(\lmb )$ is assumed to be a smooth function on the positive
axis, $G^*(s)$ is analytic everywhere except at poles.
Since we start with the assumption that there is no phase transition,
therefore the second singularity of the perturbative $G(\lmb )$
is assumed to be negative.

$G(\lmb )$ is determined by  $G^*(s)$ through the inverse Mellin 
transform 
\eqn\invmell{G(\lmb )={1\over 2\pi i}\int_C ds(2\lmb)^{-s}G^*(s),}
where the contour $C$ is a straight line with a constant
$\Re s$ satisfying $0<\Re s<3/2$, we also
rescaled $G^*(s)$ by a factor $2^{-s}$. According to
the discussion in the last section, $G^*(s)$ must have poles at all
integers and half integers with three exceptions $s=\pm 1/2$,
$s=1$. For a small $\lmb$, the contour is closed on the left
half plane, picking up poles at negative integers and negative
half integers. Since the small $\lambda$ series is convergent for
a sufficient small $\lmb$, thus the function $G^*(s)$ must grow
at most as an exponential  function for large negative $\Re s$. On the
other hand, there is no such constraint for a large positive
$\Re s$. Thus, for a large enough $\lmb$, the contour can not be
closed on the right half plane. However, one can shift the contour
rightward, thus gradually picks up negative powers in $\lmb$, and
eventually the expansion could be only an asymptotic expansion.

Thus, $G^*(s)$ must contain a factor $(s^2-1/4)(s-1)/\sin (2\pi s)$.
This factor contains all the desired poles. As pointed out earlier,
at a positive power, the residue has the leading term proportional to
$\zeta (2s)$. This factor appears for $s=n+3/2$.
However, for a large negative $\Re s$, this function
grows faster than an exponential  function. This can be seen using 
the following functional relation
\eqn\func{\zeta (2s)= 2^{2s}\pi^{2s-1}\Gamma (1-2s)\zeta (1-2s)
\sin (\pi s).}
One can not get rid of the factor $\Gamma (1-2s)$ by simply multiplying
$\zeta (2s)$ by $1/\Gamma (1-2s)$, since this factor has zeros
at positive integers and half integers. Without loss of generality,
a factor $\Gamma (a)/\Gamma (a-2s)$ can be included, where $a$ is not
an integer.
Thus one can write
\eqn\gstar{G^*(s)={(s^2-1/4)(s-1)\over \sin (2\pi s)}\left({\Gamma (a)
\zeta (2s)
\over \Gamma (a-2s)}f(s)+\dots\right),}
where $f(s)$ is not specified yet. Now, $\zeta (2s)$ vanishes if $s$
is a negative integer, thus all the poles of the form $s=-n<0$
are canceled. We shall argue later that the poles at $s=-n-1/2$
should not be canceled by zeros in $f(s)$.

For a fixed $\Re s$ and large $\Im s$, we have
$$|{1\over \sin (2\pi s)\Gamma (a-2s)}|\sim e^{-\pi |\Im s|},$$
where we have ignored a power factor. The factor $|\zeta (2s)|$
is bounded on the contour $C$. Thus, for the integral \invmell\
with the integrand given in \gstar\ to make sense, $|f(s)|$ can grow
no faster than the exponential $\exp(\pi |\Im s|)$ for large
$|\Im s |$. Also, in order to close the contour on the left half plane
to reproduce the perturbation series, $|f(s)|$ should not grow faster
than an exponential function for a large negative $\Re s$. 

Now the inverse Mellin transform dictates an asymptotic expansion for
a large $\lmb$:
\eqn\asymp{\eqalign{G(\lmb )&=-{\sin(\pi a)\Gamma (a)\over 16\pi^2}
\sum_{n=0}^\infty (n+1)(n+2)(n+4)\Gamma (n+4-a)f({n\over 2}+{3\over
2})\cr
&\zeta (n+3)(2\lmb)^{-n/2-3/2} +\dots,}}
where we have the desired factor $\zeta (n+3)$.  For $n=0$, the other 
factors should comprise to give $45/32$.

$\zeta (2s)$ vanishes if $s$ is a negative integer, as seen in \func.
It has a simple pole at $s=1/2$. This is a undesired feature, thus
$f(1/2)=0$ in order not to have a term $\lmb^{-1/2}$ in the large
$\lmb$ expansion. Although $\zeta (2s)$ does not give rise to positive
integral powers, it gives rise to positive half-integral powers
as well as to the constant term
\eqn\smal{\eqalign{G(\lmb )&=-{1\over 16\pi}f(0)+\sum_{n=1}^\infty (-1)^{n+1}
(n^2+n)(2n+3)
{\zeta (2n+2)f(-n-{1\over 2})\over 4\pi^2}\cr
&{\Gamma (a)\Gamma (2n+2)
\over \Gamma (2n+1+a)}({\lmb\over 2\pi^2})^{n+1/2}.}}
This series is convergent, since $f(-n-1/2)$ by definition does not
grow faster than an exponential function. The general perturbation 
theory indicates
that all the $\pi$ factor is what already included in $(\lmb /2
\pi^2)^{n+1/2}$, and the zeta function $\zeta (2n+2)$ is
$\pi^{2n+2}$ multiplied
by a rational number, thus $f(-n-1/2)\sim \pi^{-2n}$. If so, since
for large $n$, $\zeta (2n+2)\rightarrow 1$, the convergent
radius does not contain a factor $\pi^2$.
Indeed, if $f(-n-1/2)$ has the same sign for sufficiently large $n$,
the singular point $\lmb_c$ will be negative. Notice that the constant
term must be a rational number ($1/4$), this requires $f(0)\sim \pi$,
and this is consistent with the pattern $f(-n-1/2)\sim \pi^{-2n}$.

However, the coefficients in \asymp\ must be rational, except for the
factor $\zeta (n+3)$. It is $45/32$ at $n=0$ as computed in \gkt. It is 
fairly easy to see from the strategy of that calculation that
one will get a rational number at the order $(\ap)^{n+3}$, for each
factor $\zeta (n+3)$, ${\bf C}\zeta(n+2)$ etc. Now the factor
$\sin (\pi a)\Gamma (a)\Gamma (n+4-a)$ in \asymp\ is a rational number
times $\pi$. Thus, $f((n+3)/2)$ must be a rational number times $\pi$
to cancel the factor $\pi^2$ in the denominator in \asymp.
The only way to solve this problem is to use 
\eqn\funct{f(s)=\pi^{2s+1}h_1(s)+\pi h_2(s),}
such that $h_1(s)$ has zeros at $s=(n+3)/2$, while $h_2(s)$ has
zeros at $s=-(n+3/2)$. Both $h_i(s)$ must be bounded by exponential
functions for a general $s$. It is quite easy to construct functions
with the these properties. 

It appears therefore that it is possible to interpolate between
the small $\lmb$ expansion and the large $\lmb $ expansion. Now we
prove that the form as given in \funct\ makes the contour integral
in \invmell\ diverge. We need to show that $|h_1(s)|$ grows
as fast as $\exp (2\pi |\Im s|)$ for large $|\Im s|$. To see this, consider
the contrary, that $h_1(s)$ grows more slowly such that the following
integral
\eqn\test{{1\over 2\pi i}\int_C ds\lmb^{-s} {(s^2-1/4)(s-1)h_1(s)
\over \sin (2\pi s)},}
is well-defined, where the contour $C$ is the same as
in \invmell. Since $h_1(s)$ is bounded by an exponential function
for a large negative $\Re s$,
the above integral results in a small $\lmb$ expansion with a 
negative singularity in $\lmb$. This means that the above integral
can be extended for all positive $\lmb$, and $\lmb =\infty$ should be no
more than a regular singularity such as a branch point. On other hand,
there is no large $\lmb$ expansion resulting from integral
\test, since the integrand has no poles to the right of the contour $C$.
This implies $\lmb =\infty$ must be an essential singularity,
namely the integral for a large $\lmb$ must be smaller than any
negative power of $\lmb$. This is a contradiction. We conclude
therefore that the integral \test\ is not well-defined, and
$|h_1(s)|$ grows no more slowly than $\exp(2\pi |\Im s|)$, so
the original integral $\invmell$ is not well-defined too.

To complete our proof of the impossibility of an interpolation,
we need to argue that the form in \gstar\ is logically unavoidable,
namely, the large $\lmb$ expansion \asymp\ necessarily implies the
small $\lmb$ expansion \smal. One might think, for instance, to
have poles located on the the right half-plane only. In this case
there is no corresponding small $\lmb$ expansion, namely closing
the contour formally on the left half-plane will result in null
answer. This indicates that $\lmb =0$ is an essential singularity,
contradicting our assumption that it is just a branch point. For
example, one might start with
$$G^*(s)=\Gamma (3/2-s)\zeta (2s) f(s),$$
where the first factor has poles located at $s=(n+3)/2$. This results in
a large $\lmb$ expansion 
$$G(\lmb )=\sum_{n\ge 0}{(-1)^n\over n!}\zeta (2n+3)f(n+3/2)
(2\lmb )^{-n-3/2}.$$
If $f$ behaves as a power function asymptotically, then the above sum
is convergent for all $\lmb$, and $G(\lmb)$ is dominated by  a 
factor $\exp (-a/\lmb)$ for a small $\lmb$, and apparently 
this implies that $\lmb =0$ is an essential singularity. To remove
this factor, one need to require, for instance, $f(n+3/2)\sim
n!$. Such a factor will introduce poles on the left-half plane,
and these poles must be the ones exhibited in \gstar.

Having shown that it is impossible to extrapolate from the
small coupling regime to the large coupling regime, there is no
need to discuss how to generate terms $\lmb^n$ in the perturbation
series. Actually, if we start with a function such as $\zeta (3)
\zeta (2s-3)$, we will be able to generate the positive integral
powers. The above discussion can be repeated, and we will find
again the same type of contradiction. We also tried to introduce
poles at $s=-n$ in $f(s)$ in \gstar. It turned out at $\lmb^n$
a factor $\zeta (2n+1)$ appear. This is not permissible. In
particular, a factor $\zeta (3)$ should not appear in the two
loop calculation.

We have assumed terms
such as $\lmb^2\ln\lmb$ are absent. These terms can be included
into our consideration by introducing double poles in $G^*(s)$.

\newsec{Absence of world-sheet instantons}

Although we have provided a strong argument for the existence
of a large N phase transition somewhere in the positive axis
of $\lmb$, we are still interested in how the general large
$\lmb$ expansion series looks like. For instance, we would like to
ask whether this expansion is an asymptotic one with the
general coefficient growing as $n!$. The answer is likely negative.
One way to see this is through a more careful execution 
of the type of investigation presented in sect.2 above \rieman.
Here we provide another piece of evidence for this, the absence
of the world-sheet instantons.

Unlike in the flat background, the free energy starts to receive
contributions from the sphere in the black hole background.
The reason for this is quite simple, that the Euclidean time circle
is contractible in the black hole background, so the sphere can be
wound around this circle. Thus, the leading contribution to
the free energy is proportional to $1/g_s^2\sim N^2$.

If the general coefficient of the term $\lmb^{-(n+3)/2}$ in the
free energy goes as $n!$, we would expect that the origin of this
behavior is the existence of a spheric world-sheet instanton. For instance,
a term $\exp(-c\sqrt{\lmb})$ might be generated by a spheric minimal
surface.

The topology of the Euclidean anti-de Sitter black hole is
$S^5\times R^3\times D$, where the 5 sphere of constant curvature
is irrelevant for our question. The factor $R^3$ corresponds to the
spatial volume of D3-branes, and the infinite disk $D$ is parametrized
by the radial coordinate of the anti-de Sitter space and the Euclidean
time circle. The metric on $R^3\times D$ reads
\eqn\ebhm{ds^2={1\over R^2U^2}(U^4-U_T^4)dt^2+R^2U^2(U^4-U_T^4)^{-1}
dU^2 +{U^2\over R^2}ds_3^2,}
where $R^2=\sqrt{2\lmb}$, and $ds_3^2$ is the Euclidean metric on $R^3$.
We have set $\ap =1$. 
Using the following coordinates transformation
\eqn\trans{U^4=U_T^4(1+b^2), \quad t={R^2\over 2U_T}\psi,}
the metric is put into a simpler form
\eqn\simm{ds^2={R^2\over 4}\left({db^2\over 1+b^2}+{b^2d\psi^2\over
\sqrt{1+b^2}}+\sqrt{1+b^2}ds_3^2\right),}
where we have rescaled the metric on $R^3$. It becomes clear
that the circumference of $\psi$ is $2\pi$ in order to get
rid of a conical singularity at $b=0$. Let $b=\sinh\rho$, the
metric reads
\eqn\msim{ds^2={R^2\over 4}\left( d\rho^2 +\sinh\rho\tanh\rho d\psi^2
+\cosh\rho ds_3^2\right).}

Due to the trivial topology of $R^3$, the only reasonable possibility
for a world-sheet instanton is to embed the sphere into the coordinates
$(\rho, \psi)$. The space of this part looks like a cigar near
the tip $\rho=0$, since $\sinh\rho\tanh\rho =\rho^2-\rho^4/6 +\dots$. 
But the
actual radius of the circle gets ever larger with large $\rho$. Mapping
the sphere onto the tip of the cigar, the sphere must be at least doubly
folded. Intuitively, there can be no minimal area embedding, due to
the fact that $\sinh\rho\tanh\rho$ is an increasing function.
One can rule out this possible minimal area embedding by simply considering
the most symmetric embedding. Let the world-sheet metric be
$(dr^2+r^2d\phi^2)/(1+r^2)^2$. Without loss generality, assume 
half of the sphere $r=(0,1)$ is embedded into the tip of the cigar,
and the other half is folded back.
Let $\psi=\phi$. The boundary conditions are $\rho(r=0)=0$,
$\p_r\rho (r=1)=0$. The latter condition is to ensure the embedding be 
smooth.  The world sheet action is
\eqn\wsac{S=c(2\lmb )^{1/2}\int rdr\left((\rho ')^2 +\sinh\rho\tanh\rho
\right),}
where $c$ is a numerical constant. Use $r=\exp(-R)$, the action becomes
\eqn\wsact{S=c(2\lmb )^{1/2}\int dR\left((\rho')^2+\sinh\rho\tanh\rho
\right).}
The general solution to the minimal area problem is given by
\eqn\gens{\rho'=\pm\sqrt{C+\sinh\rho\tanh\rho},}
where $C$ is an integration constant. We can see that the boundary 
condition can not be satisfied by a nontrivial solution. The condition
$\rho (r=0)=0$ becomes $\rho ( R=\infty )=0$. This is possible only when
$C=0$. Thus the solution is
$$\rho' =\pm \sqrt{\sinh\rho\tanh\rho}.$$
The other condition, $\rho'(R=0)=0$ can never be satisfied by the above
solution unless $\rho =0$. But this is a trivial solution: The whole
sphere is mapped to the tip $\rho =0$. The same conclusion can be drawn
using the Nambu-Goto action.

We also checked that the corrected metric at the order $(\ap)^3$ as
given in \gkt\ does not allow the simplest type of world-sheet
instanton as discussed above.

\newsec{Conclusion}

The conclusion of this note is quite discouraging, that there
exists a phase transition in the large N finite temperature theory
of ${\cal N}=4$ SYM. Note that this phase transition is different 
from that discussed in \ewi, there the existence of a phase
transition has to do with the finite volume. Here the phase
transition exists for any finite temperature, although there is
evidence that there is no such phase transition at zero temperature
\dl. The continuum theory of SYM must be well-defined in the
strong coupling regime, and Maldacena's conjecture provides an efficient
way to control the theory at zero temperature in this regime.
A finite temperature theory in the strong coupling regime must be
well-defined too, and hopefully Maldacena's conjecture can
be extended to this case, as advocated in \ewi. If so, the string
theory on AdS space must exhibit the same phase transition
at a finite temperature. The transition occurs when the size
of AdS is comparable to the string scale, while the string coupling 
is kept very small.

Our result casts doubt on the program of studying the large N pure
Yang-Mills theory using Maldacena's conjecture. It is true
that the strong coupling regime shares many qualitative features
with the weak coupling regime \refs{\ewi, \elect, \miao,
\go}, such as confinement, magnetic
screening. Nevertheless to gain any quantitative control one must
study string theory in the background of the anti-de Sitter 
black hole with a small size $\lmb$. Some preliminary evidence
for disagreement between the two regimes was already noticed in \gol.

\noindent{\bf Acknowledgments} 
We would like to thank M. Douglas for collaboration on \dl.
We have benefited from conversations with A. Klemm, V. Sahakian
and R. Siebelink.
This work was supported by DOE grant DE-FG02-90ER-40560 and NSF grant
PHY 91-23780.

\listrefs
\end